\documentclass[12pt,preprint]{aastex}
\usepackage{psfig}
\newcommand{\lta}{$\; \buildrel < \over \sim \;$}
\newcommand{\simlt}{\lower.5ex\hbox{\lta}}
\newcommand{\gta}{$\; \buildrel > \over \sim \;$}
\newcommand{\simgt}{\lower.5ex\hbox{\gta}}
\newcommand{\cm}{{\rm\,cm}}
\newcommand{\kms}{{\rm\,km\,s^{-1}}}
\newcommand{\msun}{{\,M_\odot}}

\newcommand{\pb}{\phi_{\scriptscriptstyle 98}}
\newcommand{\ps}{\phi_{\scriptscriptstyle 67}} 

\newcommand{\fluxa}{{\rm\,ergs\,s^{-1}\,cm^{-2}\,\AA^{-1}}}
\newcommand{\flux}{{\rm\,ergs\,s^{-1}\,cm^{-2}}}
\newcommand{\mwd}{{\,M_{WD}}}
\newcommand{\rwd}{{\,R_{\rm WD}}}
\def \ll{\lambda\lambda}  

\slugcomment{Accepted to ApJ}
\shortauthors{Belle et al.}
\shorttitle{STIS spectroscopy of EX Hya}

\begin{document}

\title{{\it HST/STIS} Spectroscopy of the Intermediate Polar EX
Hydrae\footnote{Based on observations made with the NASA/ESA Hubble
Space Telescope, obtained at the Space Telescope Science Institute,
which is operated by the Association of Universities for Research in
Astronomy, Inc., under NASA contract NAS 5-26555. These observations
are associated with proposal 8807.}}

\author{Kunegunda E. Belle\altaffilmark{2},  
Steve B. Howell\altaffilmark{3}, Edward M. Sion\altaffilmark{4},
Knox S. Long\altaffilmark{5}, Paula Szkody\altaffilmark{6}}

\altaffiltext{2}{Astrophysics Group, Planetary Science Institute, 620
N. 6th Ave.  Tucson, AZ 85705 and X-2, MS P225, Los Alamos National 
Laboratory, Los Alamos, NM 87545; keb@psi.edu}
\altaffiltext{3}{Astrophysics Group, Planetary Science Institute, 620
N. 6th Ave.  Tucson, AZ 85705, and Institute for Geophysics and Planetary 
Physics, University of California, Riverside, CA 92521; howell@psi.edu}
\altaffiltext{4}{Department of Astronomy \& Astrophysics,
Villanova University, Villanova, PA 19085; edward.sion@villanova.edu }
\altaffiltext{5}{Space Telescope Science Institute, 3700 San Martin
Drive, Baltimore, MD 21218; long@stsci.edu} 
\altaffiltext{6}{Department of Astronomy, Box 351580, University of 
Washington, Seattle, WA 98195; szkody@astro.washington.edu}

\begin{abstract}
We present the results of our analysis of six orbits of {\it HST/STIS}
time-tagged spectroscopy of the intermediate polar, EX Hydrae.  The
high time and wavelength resolution of the {\it HST/STIS} spectra
provided an excellent opportunity to study the UV properties of EX
Hya.  Measurements of the continuum and emission line fluxes
corroborate earlier studies that show that the emission line fluxes are
modulated more strongly than the continuum flux and originate from the
accretion curtains, while the continuum flux originates from the white
dwarf photosphere.  The measured $K$ amplitude of the narrow emission
line radial velocity curve is used to calculate a white dwarf mass
of $0.91\pm0.05\msun$.  Synthetic white dwarf photosphere
and accretion disk spectral models are used to further refine the white
dwarf and accretion disk properties.  Using the spectral models, it is
determined that EX Hya has a white dwarf of mass $0.9\msun$, $T=23000$
K, an accretion disk truncated at $2.5\rwd$, and is at a distance of
$60$~pc.
\end{abstract}

\section{INTRODUCTION}

Intermediate polars (IPs) are a class of magnetic cataclysmic
variables in which a white dwarf primary star sustains a magnetic
field that is sufficient enough to divert the gravitationally
controlled mass transfer.  IPs may have a truncated accretion disk
(the inner edge being wiped out by the magnetic field) or may be
diskless accretors (accreting directly from the ballistic accretion
stream), but the defining characteristic that separates IPs from the
higher magnetic field polars is an asynchronously rotating white
dwarf.  As the magnetic field lines of the white dwarf sweep around
the inner edge of the accretion disk, curtains of magnetically
controlled material form and follow along the field lines until the
material impacts the white dwarf surface at or near the magnetic
poles, with both upper and lower poles in IPs accreting equally.

EX Hydrae is an IP with a spin period of 67 minutes and an orbital
period of 98 minutes, which causes an unusual scenario in which the
corotation radius is close to the $L_1$ point \citep{king99}.
Magnetically controlled accretion occurs from the inner edge of the
disk, and possibly from the outer edge of the disk, as evidenced by
dips seen during the bulge dip in EUV light curves of EX Hya
\citep{bel02}.  Radiation from EX Hya is seen to be modulated on both
the binary orbital and white dwarf spin periods.  Binary modulation is
seen in the form of a large bulge dip at phases $\pb=0.5-1.1$ ($\pb$
refers to the binary phase) in the EUV light curve, which is caused by
absorption along the line of sight to the emitting poles due to the
extended hot spot on the edge of the accretion disk.  Continuum
emission is modulated sinusoidally over spin phase for all wavelength
regimes \citep[e.g.,][]{hellier87,rosen91,bel02}.  Maximum spin phase
occurs when the upper accretion pole points away from the observer,
allowing a direct view of upper and lower accretion regions; minimum
spin flux occurs when the upper pole points towards the observer and
the lower pole is self-eclipsed by the white dwarf.

Previous UV observations of EX Hya consist of HUT spectra
\citep{gre97}, {\it ORFEUS II} and {\it IUE} spectra \citep{mauche99},
and {\it HST/FOS} spectra \citep{eis02}.  These observations show that
the broad emission lines and the continuum flux are formed in
different regions; the emission lines from an optically thick
accretion curtain and the continuum from the inner most part of the
accretion curtain and the white dwarf itself.  Blue-shifted absorption
was also seen in the \ion{C}{4} $\lambda1549$\AA\ emission line.  The
UV data of EX Hya have also been modeled using a variety of continuum
emission sources.  The continuum has been fit by a $T=25000$ K
blackbody \citep{gre97}, but this same work also showed that a
bremsstrahlung model, with $T=74000$ K, or a white dwarf photosphere
of $T=26200$ K also fit the data well.  \citet{eis02} found that a
white dwarf photosphere of $T_{\rm eff}=25000\pm3000$ K (determined
from line ratios) fit their {\it FOS} spectra well.

\section{OBSERVATIONS}
Our {\it HST/STIS} observations, obtained through DDT time, comprised
six orbits and used the E140M grating in Echelle mode.  Each orbit
produced a time-tagged spectrum consisting of 40 Echelle orders.  The
E140M grating spans a wavelength range of $1150-1735$\AA\ and has a
resolving power of $R=45,800$, which corresponds to a wavelength
resolution of $0.013-0.019$\AA/pixel throughout the spectrum.  The six
orbit observation provided complete coverage of EX Hya's spin orbit,
but only $\sim2/3$ of its binary orbit.  Table \ref{phase} gives the
specific phase coverage of each data set.  Each spectrum was processed
with IRAF/STSDAS software and the time-tagged data was utilized to bin
the spectra into phase resolved components.  

\begin{figure}
\plotone{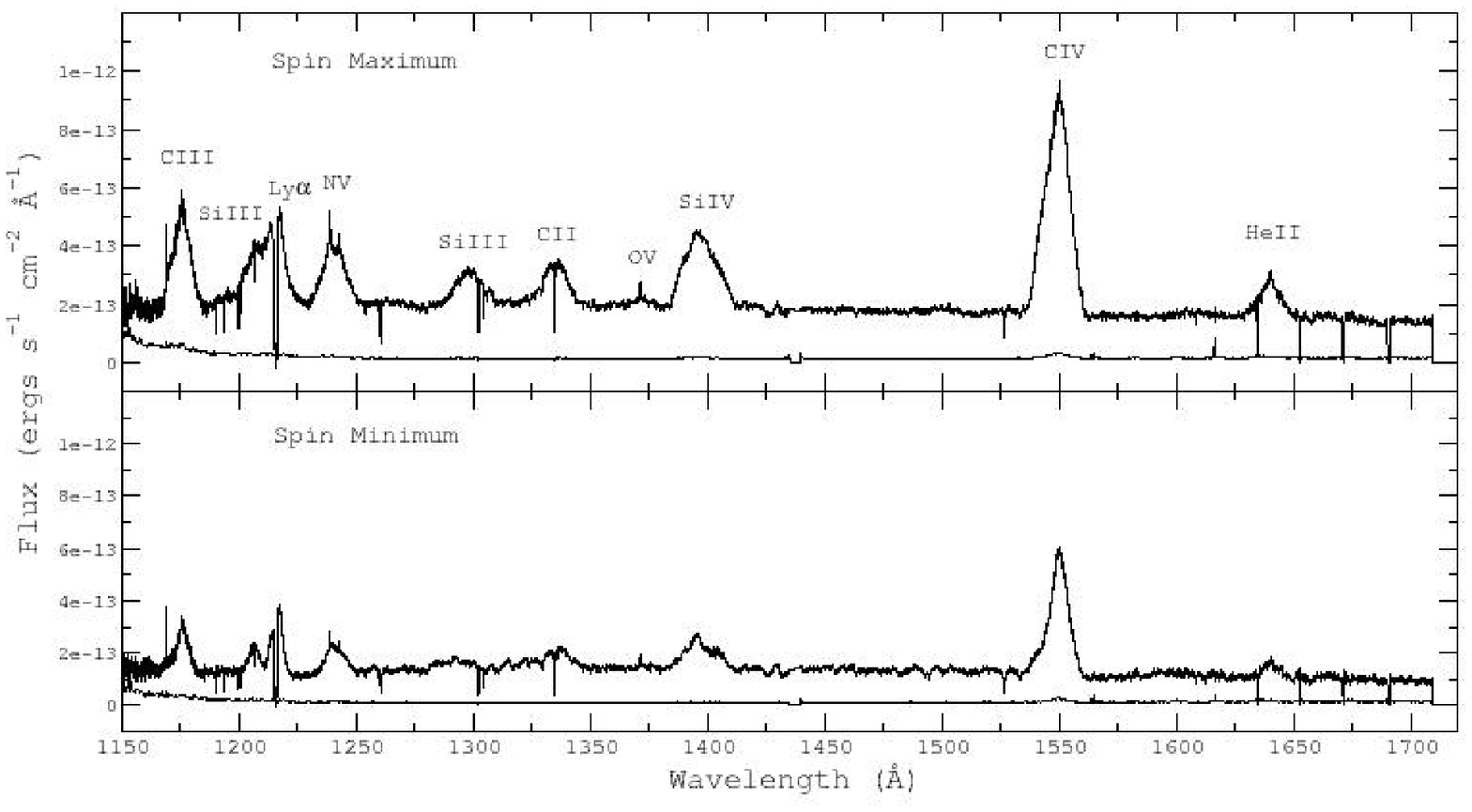}
\figcaption[]{Our {\it HST/STIS} data set separated into spin
maximum and spin minimum phases.  The spectra contain broad emission
lines of \ion{C}{3} $\lambda1175$\AA, \ion{Si}{3} $\lambda1206$\AA,
$\lambda1298$\AA, Ly$\alpha$ $\lambda1216$\AA, \ion{N}{5}
$\lambda1240$\AA, \ion{C}{2} $\lambda1335$\AA, \ion{O}{5}
$\lambda1371$\AA, \ion{Si}{4} $\lambda1396$\AA, \ion{C}{4}
$\lambda1549$\AA, and \ion{He}{2} $\lambda1640$\AA, along with narrow
emission lines \ion{N}{5} $\ll1238,1242$\AA\ and \ion{O}{5}
$\lambda1371$\AA.  The thin line in both plots represents the
$1\sigma$ errors. \label{hst_spec}}
\end{figure}

Spectra were extracted for a variety of orbital and spin phase bins.
Spin phase bins were determined according to the spin modulation seen
in the EUV light curve of EX Hya \citep{bel02} and we have defined
spin maximum phases as $\ps=0.865-0.365$ and spin minimum phases as
$\ps=0.365-0.865$, where $\ps$ refers to the 67 minute spin ephemeris,
$T=2437699.8914(5) + 0.046546504(9)\mathrm E - 7.9(4)\times 10^{-13}
\mathrm E^2$ \citep{ephem}.  Spectra extracted throughout the spin
period were binned into $\Delta\ps=0.1=402.2$ s phase bins; bins were
chosen as such to preserve a moderate signal-to-noise ratio.  Three
types of binary phased spectra were created: binary phased, spin
maximum binary phased, and spin minimum binary phased.  For the binary
ephemeris, we used $T=2437699.94179 + 0.068233846(4)\mathrm E$
\citep{ephem}.  Spin maximum data existed for binary phases
$\pb=0.0-0.7$, but spin minimum data only existed for binary phases
$\pb=0.0-0.6$.  Each binary phase bin has a resolution of
$\Delta\pb=0.1=589.5$ s, where $\pb$ refers to the 98 minute orbital
period.  We used IRAF/SPLOT to analyze the spectra and determine the
$1\sigma$ errors we report here.  Figure \ref{hst_spec} presents our
complete spectral data set divided into two phase bins: spin maximum
and spin minimum.  The spin maximum spectrum has on overall higher
continuum level and stronger emission lines than the spin minimum
spectrum.  Each spectrum exhibits broad emission lines of \ion{C}{3}
$\lambda1175$\AA, \ion{Si}{3} $\lambda1206$\AA, $\lambda1298$\AA,
Ly$\alpha$ $\lambda1216$\AA, \ion{N}{5} $\lambda1240$\AA, \ion{C}{2}
$\lambda1335$\AA, \ion{O}{5} $\lambda1371$\AA, \ion{Si}{4}
$\lambda1396$\AA, \ion{C}{4} $\lambda1549$\AA, and \ion{He}{2}
$\lambda1640$\AA, along with narrow emission lines \ion{N}{5}
$\ll1238,1242$\AA\ and \ion{O}{5} $\lambda1371$\AA, which appear in
all phases except for the early binary phases ($\pb=0.0-0.1$).

\section{DISCUSSION AND ANALYSIS}
\subsection{Continuum} 
Continuum flux measurements were made at three emission line-free
wavelength regions along each spectrum: $\lambda\lambda1352-1362$\AA,
$\lambda\lambda1440-1520$\AA, and $\lambda\lambda1565-1620$\AA.  The
results of the $\lambda\lambda1352-1362$\AA\ measurements are shown in
Figure \ref{spcont_flux} (spin phased).  The other two wavelength
regions exhibit similar behavior.  Error bars on each measurement are
RMS values as reported by IRAF/SPLOT.  The spin phased continuum flux
(Figure \ref{spcont_flux}) is fit well by the sinusoidal function,
$f(10^{-13}\fluxa)=A+B\sin 2\pi(\ps-\phi_0)$, where $A=1.68$,
$B=0.38$, and $\phi_0=0.76\pm0.05$, which places the peak of the light
curve at $\ps=0.01\pm0.05$.  The error here is dominated by the fact
that each phase bin measurement was confined within the
$\Delta\ps=0.1$ phase bin resolution of the data.  The continuum flux
is modulated by $B/A=23\%$.  The peak in UV continuum flux is earlier
than that of the EUV emission, which occurs at $\ps=0.115\pm0.001$
\citep{bel02}.  As both the {\it HST} and {\it EUVE} observations of
EX Hya occurred simultaneously, an error in the ephemeris is ruled out
as a cause for phase difference in peak continuum fluxes; rather, the
phase disparity is likely a result of distinct EUV and UV continuum
emission sites.  The sinusoidal fits to all of the spin phased UV
continuum wavelength regions are given in Table \ref{contsinfit}.

The spectra were also measured over binary phase, and, in agreement
with previous studies \citep[e.g.,][]{hellier87}, we found that the
continuum flux did not exhibit coherent modulation over the binary
orbital period.

\subsection{Emission Lines}
\subsubsection{Broad Emission Lines}\label{bels}

The broad emission lines were fit with single Gaussian functions and a
linear continuum, using IRAF/SPLOT, to determine parameters such as
central wavelength, flux, FWHM, and equivalent width.  The Ly$\alpha$
and \ion{Si}{3} $\lambda1206$\AA\ lines were not measured because they
are blended during the spin maximum phases.

\begin{figure}
\plotone{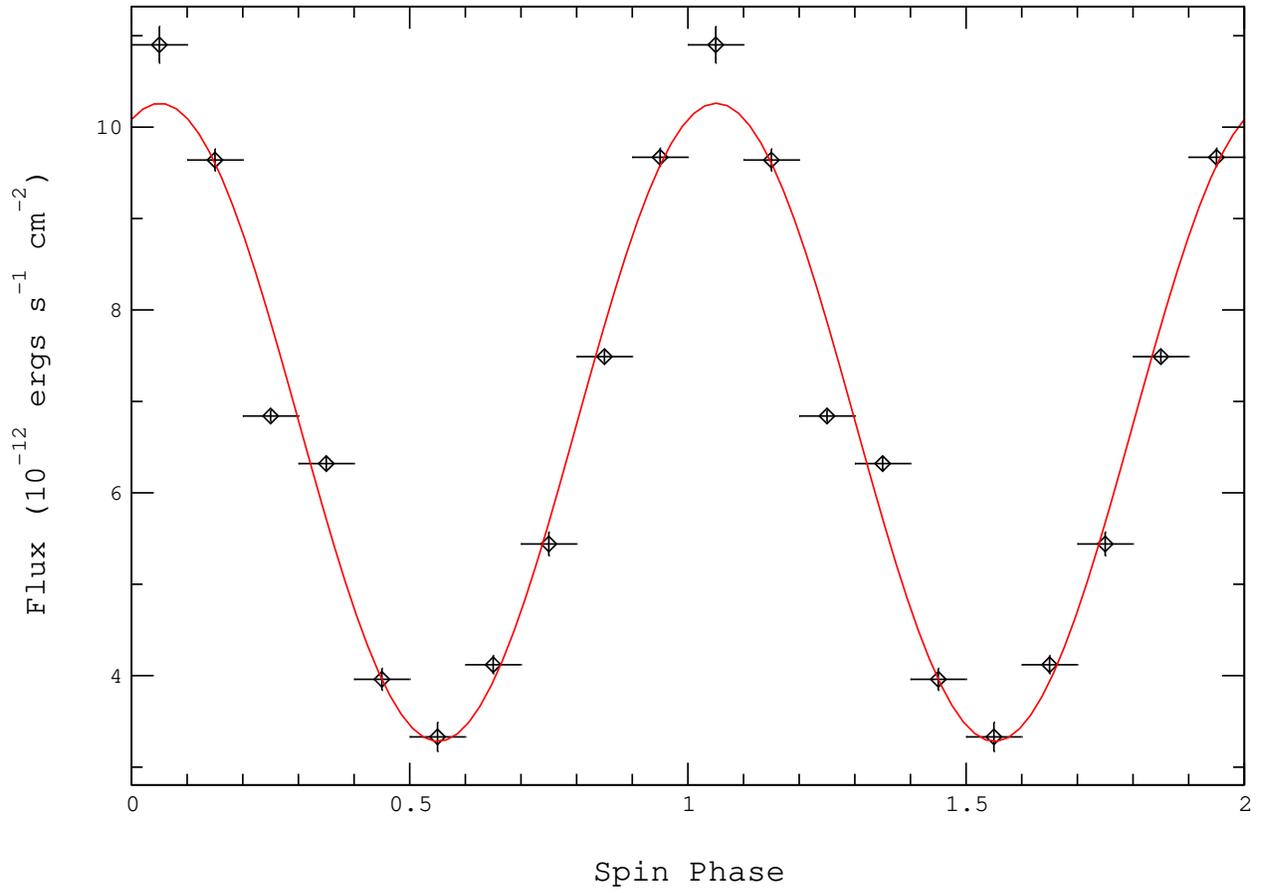}
\figcaption[]{\ion{C}{4} broad emission line flux plotted over spin
phase and fit with a sinusoidal function (solution given in Table
\ref{fluxsinfit}). \label{c4fluxsp}}
\end{figure}

Over spin phase, the emission line fluxes exhibit a sinusoidal shape
(\ion{C}{4} is shown as an example in Figure \ref{c4fluxsp}), and can
be fit well with a sinusoidal function of the form
$f(10^{-12}\flux)=A+B\sin 2\pi(\ps-\phi_0)$.  The solutions for each
line are given in Table \ref{fluxsinfit}.  The emission peaks near
$\ps\sim0.05-0.1$ for each line.  Considering the error of $0.05\ps$
on each phase bin, the emission line fluxes peak at roughly the same
phase as the continuum flux ($\ps\sim0$).  \ion{C}{3}
$\lambda1175$\AA\ has the greatest modulation, at $73\%$, but all of
the lines are strongly modulated by around $50\%$.  Compare this with
the continuum with a flux modulation of only $\sim20-25\%$.  While
both the continuum and emission lines are modulated sinusoidally over
spin phase, the difference in the modulation amplitude suggests that
the continuum and emission lines have different formation sites.  Over
binary phase the emission line flux is not modulated coherently.
Comparison of the emission line fluxes over binary and spin phase
demonstrates that the flux modulations are clearly a function of spin
phase, rather than binary phase.

\begin{figure}
\plotone{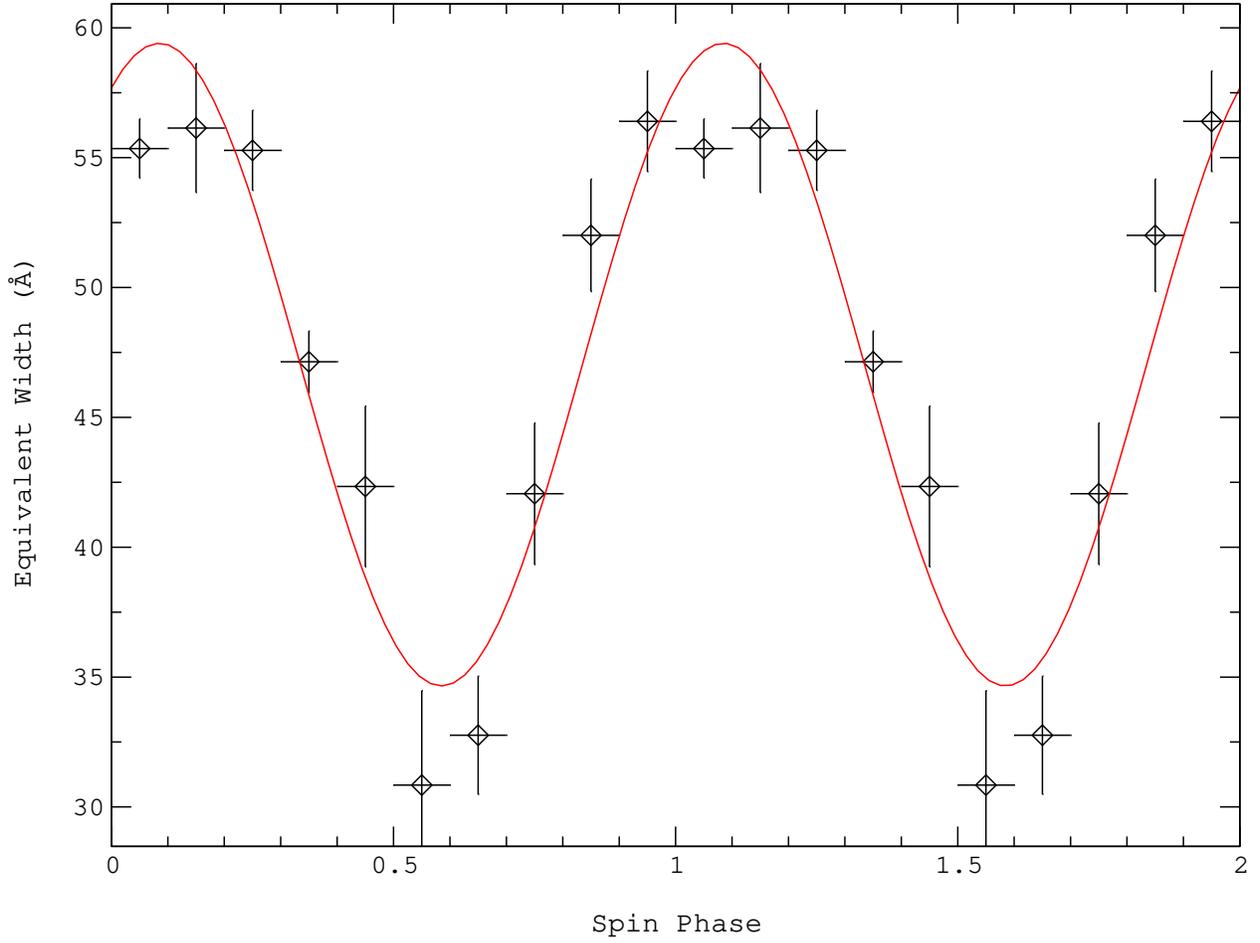}
\figcaption[]{Equivalent width of the \ion{C}{4} broad emission line phased
over the spin period.  Solutions for the sinusoidal fits to all of the
broad emission lines are given in Table \ref{eqwsinfit}. \label{eqws}}
\end{figure}

Equivalent widths (EWs) of strong UV emission lines from optically
thin plasmas are typically on order of 100s of \AA ngstroms.  The EWs
of the emission lines appearing in the EX Hya spectra are smaller than
expected for optically thin plasma emission.  \ion{C}{4}, which has
the largest EW of all the broad emission lines, is shown as an example
in Figure \ref{eqws} (solutions for the sinusoidal fits to all of the
lines are given in Table \ref{eqwsinfit}).  There are two possible
scenarios that may explain the observed EWs.  The first is that
optically thick material is responsible for the emission lines.  The
second is that the optically thin material is only weakly emitting and
another form of radiation, such as cyclotron emission, may play an
important role in the radiative cooling of the accretion curtains.  If
the latter is true, evidence should be seen in the form of cyclotron
humps; for a magnetic field strength of $B\sim5$ MG, the first
cyclotron harmonic should appear around 10$\mu$.  The almost
triangular shapes of the lines implies that they are not from a purely
optically thin material, thus it is likely that both optically thin
and optically thick emission contributes to each emission line.

\begin{figure}
\plotone{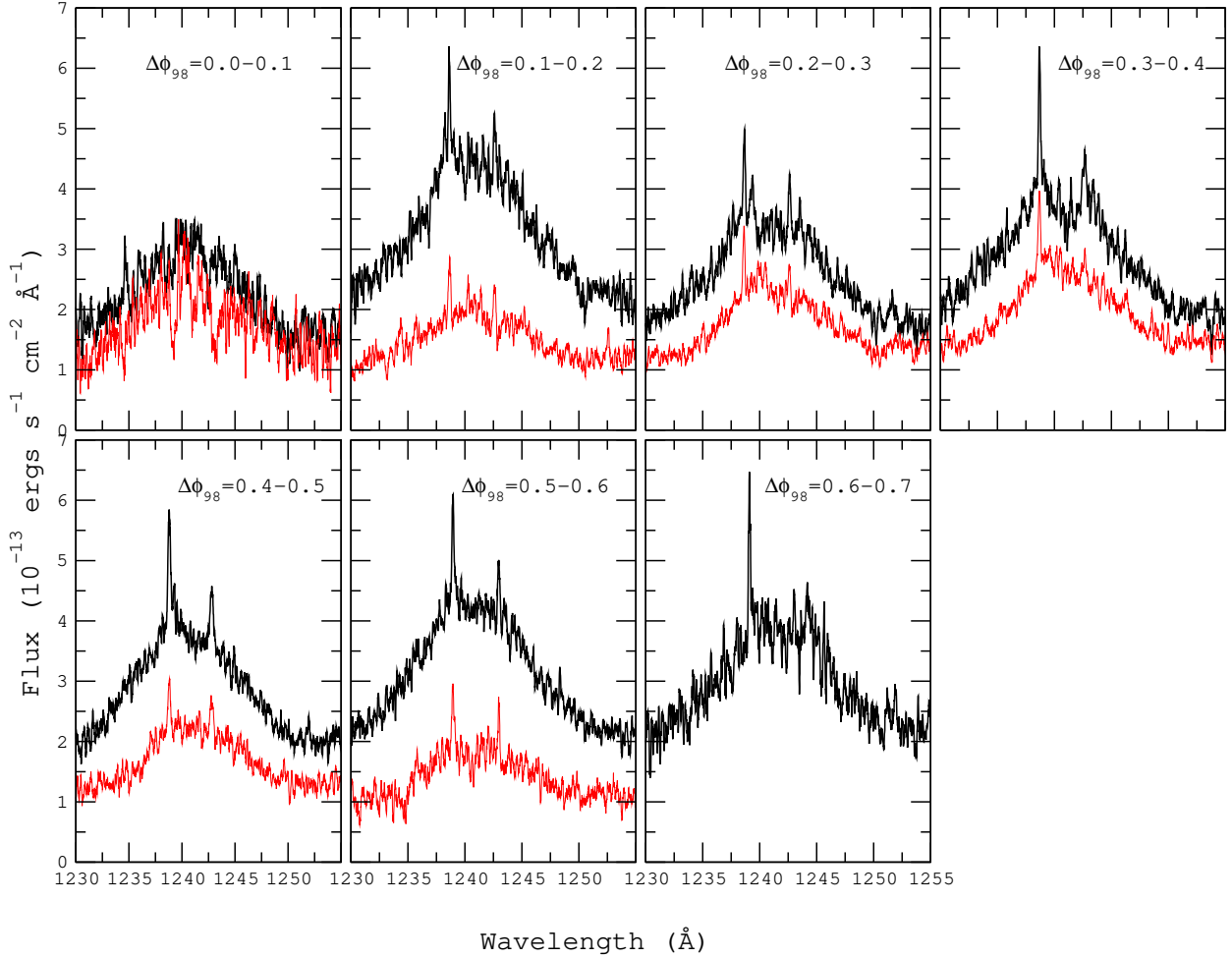}
\figcaption[]{\ion{N}{5} $\lambda1240$\AA\ emission line profile
evolution over the binary orbit and separated into spin maximum (thick
line) and spin minimum (thin line).  For all phases except
$\pb=0.0-0.1$, the line profile extracted at spin maximum is greater
in flux than the spin minimum line.  Also note the narrow emission
lines that appear during phases $\pb=0.1-0.7$ and that exhibit a
double-peaked nature in the spin maximum spectrum near binary phases
$\pb=0.1-0.2$, while the blue component only is visible in the spin
minimum line profile. \label{n5}}
\end{figure}

\begin{figure}
\plotone{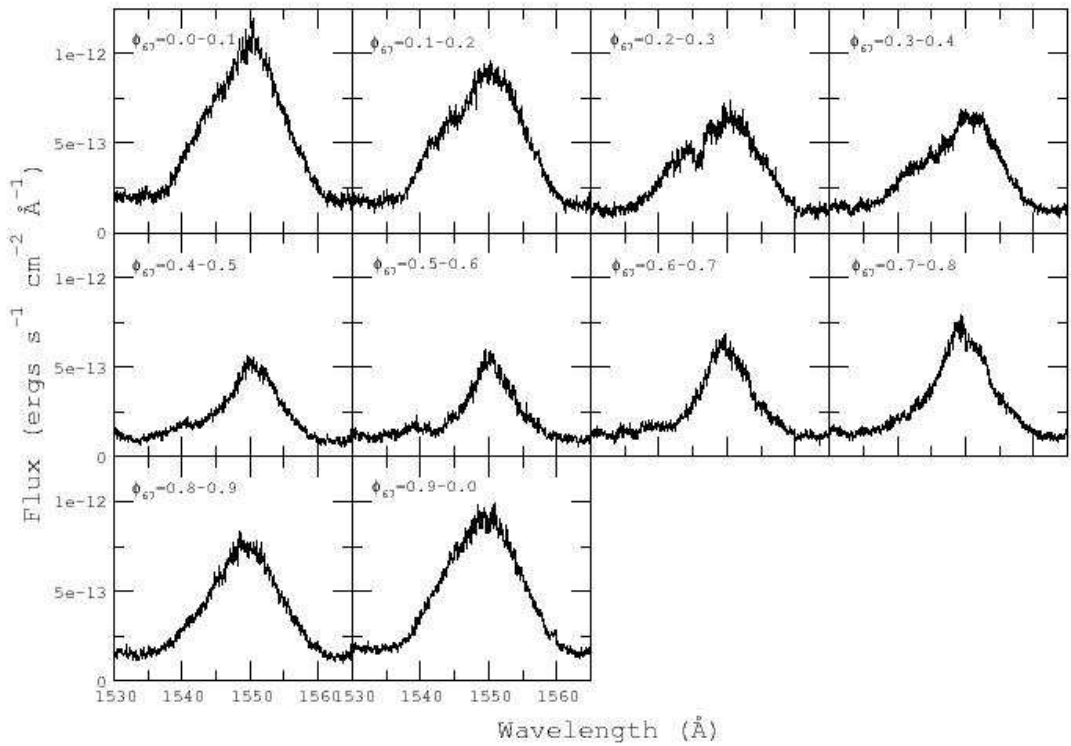}
\figcaption[]{\ion{C}{4} $\lambda1549$\AA\ line over spin phase.
Note the blue-shifted absorption that appears during phases
$\ps=0.2-0.3$ and the asymmetric profile as the line evolves over spin
phase. \label{c4sp}}
\end{figure}

Figures \ref{n5} and \ref{c4sp} present two examples of the broad
emission line evolution over binary and spin orbit.  The \ion{N}{5}
$\lambda1240$\AA\ line shown in Figure \ref{n5} has been separated
into spin maximum (thick line) and spin minimum (thin line) and phased
over the binary orbit.  The line retains roughly the same form
throughout the binary orbit for both spin phases.  Aside from binary
phase bin $\Delta\pb=0.0-0.1$, the flux of the spin maximum line is
constantly greater than the flux of the spin minimum line.  This
suggests that the lines are formed in a region that would be partially
absorbed or occulted during spin minimum phases, i.e. the inner part
of the accretion curtain.  Also of interest is the lack of narrow
emission lines during binary phases $\Delta\pb=0.0-0.1$, which
correspond to the egress of the bulge dip; absorption features appear
in their place.  The narrow lines will be discussed in further detail
in \S\ref{nels}.  The fact that the lines have roughly the same amount
of absorption during $\Delta\pb=0.0-0.1$ implies that the lines are
formed interior to the bulge.

If the change in flux of the broad emission lines between spin phases
maximum and minimum is due to absorption through the accretion
curtain, we can estimate the density of the curtains from the amount
of absorption.  Using the simple relation of $I=I_0e^{-\tau}$, we may
first deduce $\tau$ and then assume a Kramer's opacity law to
determine the density.  The physical size of the accretion curtains
must first be estimated.  At most, the radial extent of the footprint
of the curtain on the disk could be the entire truncated accretion
disk and at the least, it would only be a small fraction of the disk
size.  The disk size was determined by measuring optical emission
lines (using the optical data obtained by us during this campaign) to
determine the inner and outer accretion disk radius, assuming a
Keplerian accretion disk.  We considered two white dwarf masses (see
\S\ref{mass}), $\mwd=0.90\msun$ and $\mwd=1.33\msun$, and calculated
the disk radii corresponding to each white dwarf mass ($R_{\rm
inner}=1.6-2.3\times10^9\cm$ and $R_{\rm
outer}=2.5-5.9\times10^9\cm$).  Several different physical sizes, $S$,
of the accretion curtain were used for our density calculation.  We
take $S=aR_{\rm disk}$, where $a=0.01,0.10,0.50$ and $1.0$.  In this
way we were able to sample a range of physical depths for the
accretion curtain.  The emission line fluxes were measured during spin
maximum and spin minimum, excluding binary phases of the bulge dip.
This ensures that material associated with the bulge, which is also
responsible for absorption at certain spin phases, will not be
mistakenly identified with accretion curtain material.  Five emission
lines were used for this calculation; \ion{C}{3}, \ion{N}{5},
\ion{Si}{4}, \ion{C}{4}, and \ion{He}{2}.  \ion{Si}{3}
$\lambda1296$\AA\ and \ion{C}{2} were not used due to their ambiguous
line profiles during spin minimum.  For each white dwarf mass and
corresponding $S$ values, we find that $\tau=0.7-1.1$ and the density
ranges between $10^{-10}-10^{-9}\,{\rm g\,cm^{-3}}$ for all values of
$S$.  If we assume that the absorbing material is pure hydrogen, we
obtain a number density of $N_{\rm H}=7\times10^{13}-8\times
10^{14}\cm^{-3}$ within the accretion curtains.

The \ion{C}{4} $\lambda1549$\AA\ emission line has a blue-shifted
absorption component that appears during spin phases
$\Delta\ps\sim0.2-0.8$ as evidenced in Figure \ref{c4sp}.  A similar
feature was also observed in the \ion{C}{4} line in HUT data during
spin phases $\Delta\ps=0.36-0.48$ (their data only extended through
$\ps=0.5$, Greeley et al. 1997).  Although their \ion{C}{4} line is
phased into finer orbital bins, our \ion{C}{4} line clearly shows an
absorption component that begins prior to $\ps=0.3$ and evolves into
an absorption that eats away the blue wing of the emission line.  At
phases $\ps=0.2-0.3$, the blue absorption has a FWHM of $\sim275\kms$
and is blue-shifted by $\sim615\kms$ from the center of the \ion{C}{4}
doublet.  This is slightly less than the value of $\sim1000\kms$
reported by \citet{gre97} for their blue \ion{C}{4} absorption
feature.

\ion{C}{3}, \ion{N}{5}, and \ion{C}{2} also show a blue absorption
component during spin phases $\ps\sim0.2-0.7$.  The \ion{N}{5}
absorption component has a FWHM of $\sim400\kms$ and is blue-shifted
by $\sim700\kms$ from the center of the line.  The \ion{C}{3}
absorption component is really too shallow to accurately measure,
though it is visible during spin phases $\ps\sim0.2-0.7$.  The
\ion{C}{2} line seems to have a blue-shifted absorption component
throughout the entire spin phase.  Its FWHM ranges from $655\kms$ at
$\ps=0.4-0.5$ to $\sim200\kms$ at $\ps=0.0-0.1$ while the blue-shift
from the center of the emission line remains fairly constant at
$\sim1400\kms$.  However, the \ion{C}{2} absorption feature is
associated with a white dwarf absorption line (see \S\ref{model}),
which may influence the motion of the line.

\begin{figure}
\plotone{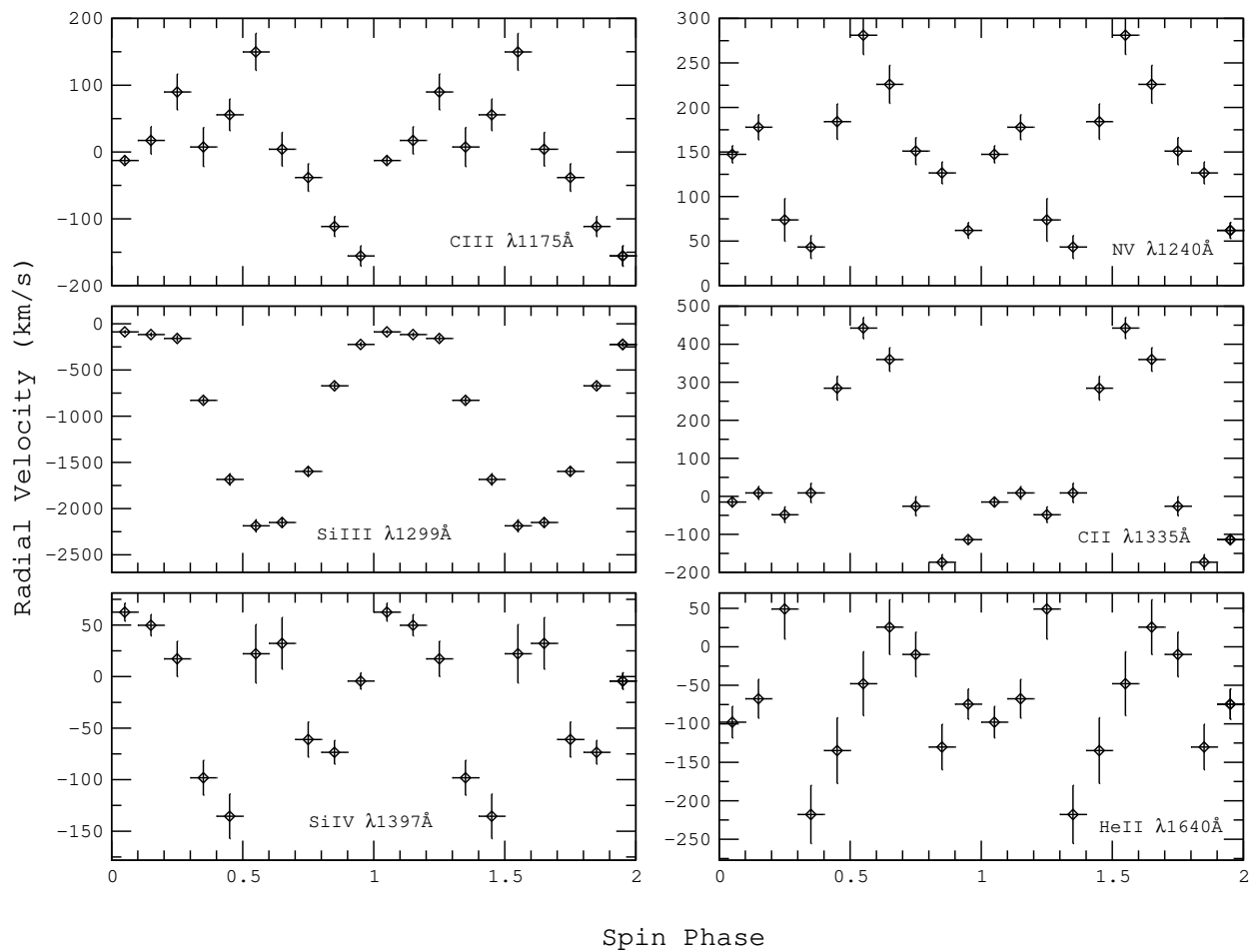}
\figcaption[]{Broad emission line radial velocities phased on the spin
period.  \ion{Si}{3} is the only line that appears to have a
sinusoidal shape; the other radial velocity curves are somewhat
double-peaked, with one peak occurring near $\ps=0.2$ and the second
occurring at $\ps=0.5$. \label{bel_rv_sp}}
\end{figure}

Radial velocity curves for the broad emission lines are shown in
Figure \ref{bel_rv_sp} phased over the spin period.  The spin phased
radial velocity curves present surprising results.  In previous EX Hya
observations \citep[e.g.,][]{hellier87}, the maximum blue-shift of the
optical emission lines observed around $\ps=0.0$ has been used to
place the origin of these lines in the accretion curtain; while the
upper pole points away from us, material flowing into the pole will be
at its maximum blue-shift.  We see that three lines exhibit this
behavior: \ion{C}{3}, \ion{N}{5}, and \ion{C}{2}. \ion{Si}{3},
\ion{Si}{4}, and \ion{He}{2}, on the other hand, have maximum
blue-shifts closer to $\ps\sim0.5$.  A double peaked structure is
visible in all but the \ion{Si}{3} radial velocity curve.

\subsubsection{Narrow Emission Lines}\label{nels} 
\begin{figure}
\plotone{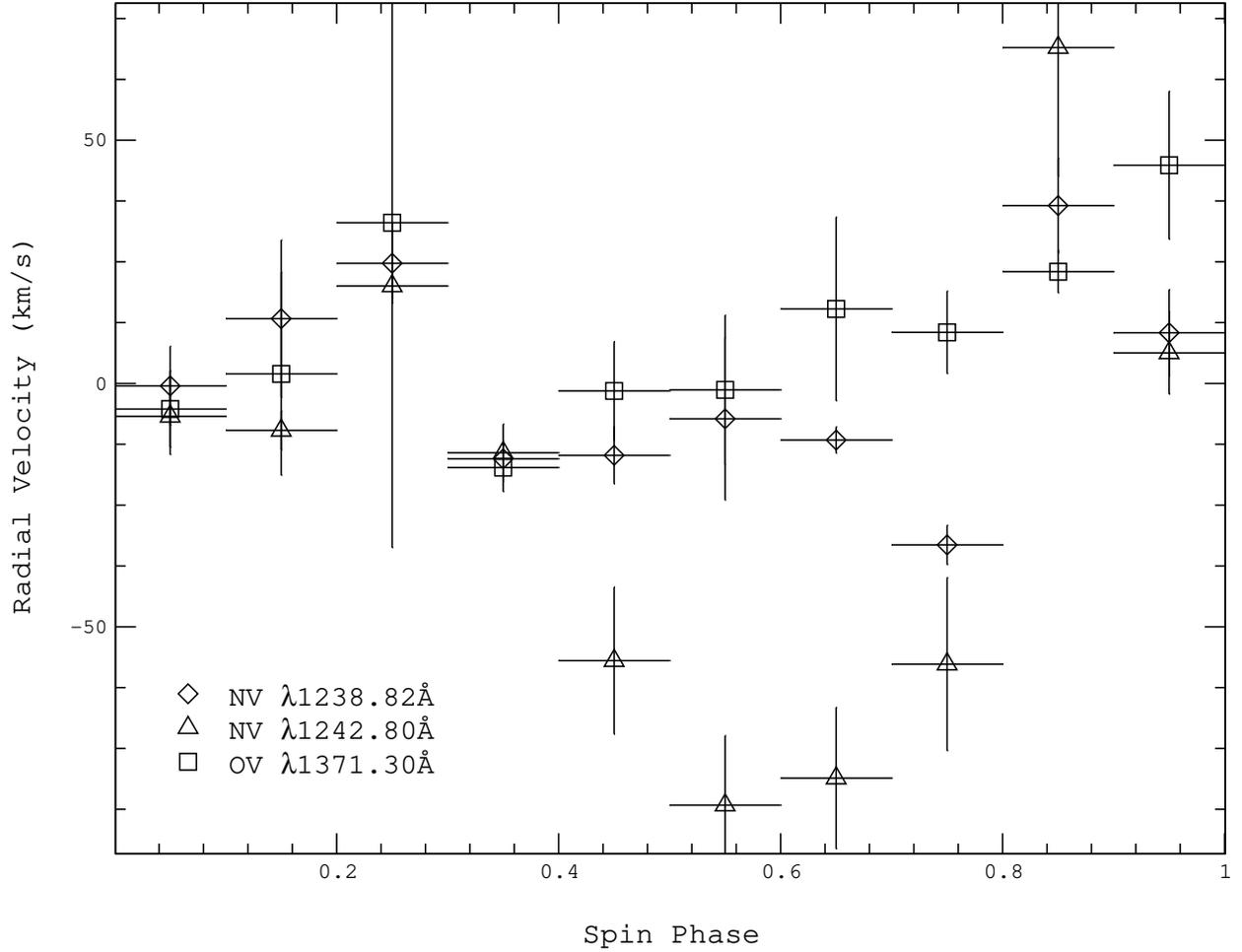} 
\figcaption[]{Narrow emission line radial velocities phased on the
spin orbit.  The radial velocity curve of each line is double-peaked,
with maxima occurring at $\ps=0.2$ and $\ps=0.8$; the \ion{N}{5}
$\lambda1242$\AA\ component shows the strongest modulation.
\label{nel_rv_sp}}
\end{figure}

Three narrow emission lines appear in the {\it HST} spectra:
\ion{N}{5} $\ll 1238.8,1242.8$\AA\ and \ion{O}{5} $\lambda1371.3$\AA,
and represent the two highest ionization energy lines present in the
spectrum, with $\chi_{\rm NeV}=98$ eV and $\chi_{\rm OV}=114$ eV.
These lines have FWHM$\sim0.2$\AA\ and are visible over both spin and
binary orbit, and only disappear during binary phases $\pb=0.0-0.1$,
which correspond to the egress of the bulge dip.  A radial velocity
curve over spin phase is shown for the narrow emission lines in Figure
\ref{nel_rv_sp}.  Each radial velocity curve has a double-peaked
structure, although the \ion{O}{5} and \ion{N}{5} $\lambda1238$\AA\
lines are less modulated than the \ion{N}{5} $\lambda1242$\AA\ line.
For each line the velocities tend to be blue for two reasons: one,
only two-thirds of the binary orbit, $\pb=0.0-0.7$, are (unevenly)
sampled in each spin phase bin and two, a $\gamma-$velocity has not
been subtracted.  Previous measurements of the $\gamma-$velocity
(derived from optical spectroscopy) range from $-9\pm35\kms$
\citep{gil82} to $-180\kms$ \citep{hellier87}, and we will show below
that we obtain a value slightly different from the \citet{gil82} and
\citet{hellier87} results.

\begin{figure}
\plotone{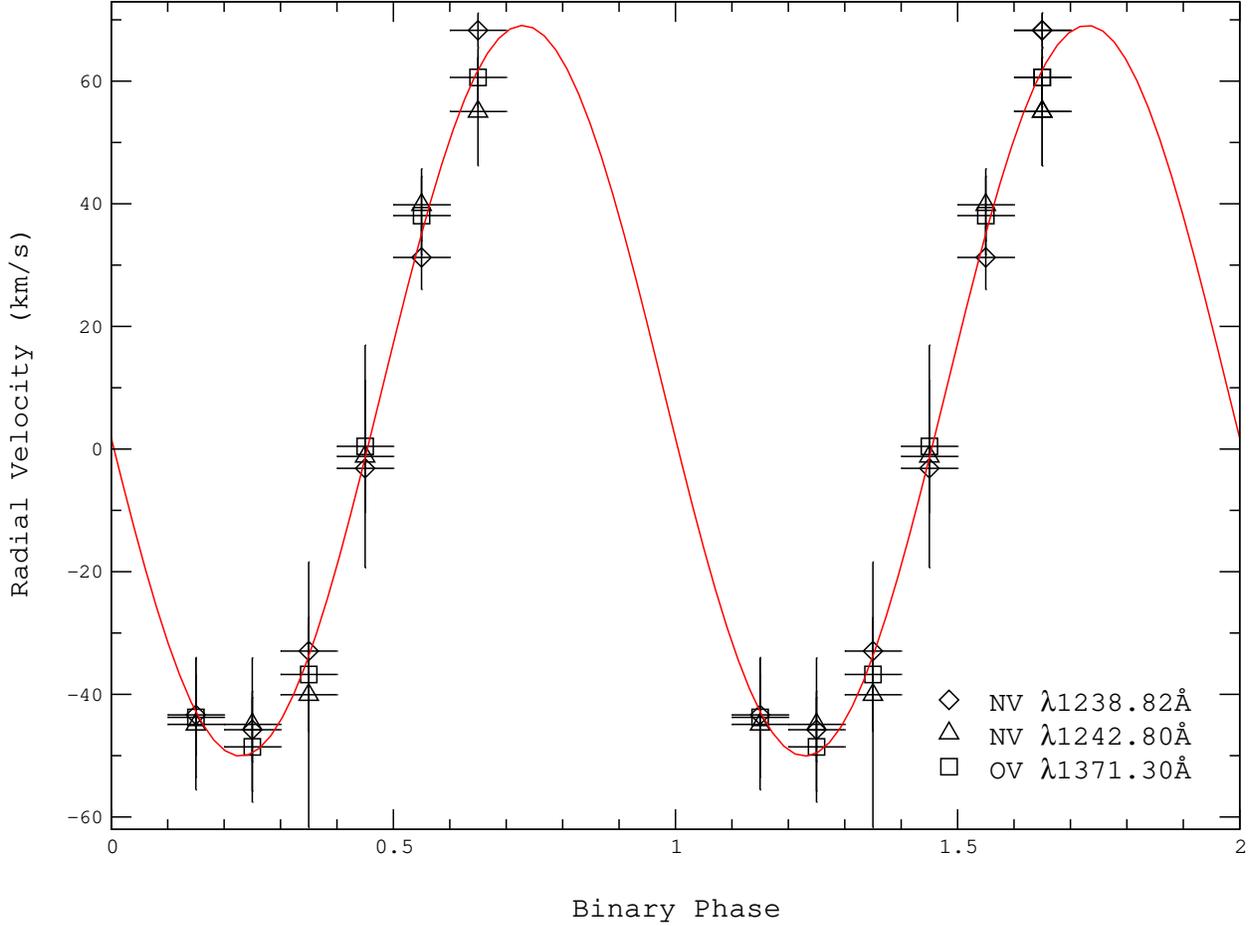}
\figcaption[]{Binary phased radial velocity curve for the narrow
emission lines.  This radial velocity curve is fit nicely with a
sinusoidal function of the form $v=\gamma + K\sin2\pi(\pb-\phi_0)$,
with $\gamma=9.5\pm3\kms$, $K=59.6\pm2.6\kms$ and
$\phi_0=0.98\pm0.05$. \label{nel_rv_bp}}
\end{figure}

The radial velocities of the narrow emission lines over the binary
phase are shown in Figure \ref{nel_rv_bp} and can be fit with a
sinusoidal function of the form $v=\gamma + K\sin2\pi(\pb-\phi_0)$,
with $\gamma=9.5\pm3\kms$, $K=59.6\pm2.6\kms$ and
$\phi_0=0.98\pm0.05$.  The $K$ amplitude derived here is comparable
with a previously determined value of $K_1=69\pm9\kms$
\citep{hellier87}.  The small velocities, the clear sinusoidal
modulation of the velocities, and the $K$ amplitude suggest that these
lines are formed near or on the white dwarf surface.  A high
temperature emission region for these lines increases the likelihood
that the lines are not contaminated by emission from the accretion
disk.  This narrow line radial velocity curve is therefore likely to
be a true representation of the motion of the white dwarf, as compared
to optical (disk) emission lines that have been used previously to
determine the $K_1$ amplitude and the $\gamma-$velocity
\citep[e.g.,][]{hellier87}.  We would like to note that the proximity
of the narrow line emission region to the white dwarf surface should
cause these lines to be gravitationally redshifted.  Unfortunately,
without an accurate $\gamma$-velocity measurement, it is impossible to
separate the gravitationally redshift from the $\gamma$-velocity of the
narrow emission line radial velocity curve.
  
The narrow emission lines appear to be double-peaked during certain
spin phases.  The \ion{N}{5} doublet shows a strong double-peaked
profile during spin phases $\ps=0.1-0.3$, $\ps=0.5-0.6$, and
$\ps=0.9-1.0$.  \ion{O}{5} is double-peaked during spin phases
$\ps=0.0-0.3$, $\ps=0.4-0.6$ and $\ps=0.7-0.8$.  These phases roughly
correspond to spin maximum phases.  Close inspection of Figure
\ref{n5} also shows the double-peaked nature of the \ion{N}{5} line
during spin maximum and that during spin minimum, it is only the blue
component of the double-peaked line that is still visible.  The
radial velocities of these lines over binary phase tell us that they
originate close to the white dwarf surface.  The double-peaked nature
of the lines suggest a two velocity component as their origin.  We
propose that these lines are formed in a wind or material outflowing
from the poles.  During spin maximum, we have a direct view of both
poles (possibly not a direct view of the northern pole as it points
away from us, but a wind would clearly be seen) giving us both red and
blue-shifted components.  At spin minimum, when the upper pole points
towards the observer, the lower pole and its (red-shifted) emission is
blocked from view; only the blue-shifted narrow line component would
be seen at these phases.  As discussed in \S\ref{bels}, evidence of a
wind is also seen in the form of blue-shifted absorption in several of
the broad emission lines.

\subsection{Mass Determinations}\label{mass}
In order to calculate a white dwarf mass, a value for the secondary
mass must be estimated.  Previous studies have used an $M_2$ value
determined from a zero age main sequence (ZAMS) mass-radius relation
applied to the Roche lobe radius for a given binary orbital period.
\citet{hellier87} found $M_2=0.128\msun$ using the ZAMS mass-radius
relation given in \citet{pat84} and an earlier study used the relation
presented in \citet{war76} to find $M_2=0.18\msun$.  We use a value
based on an up-to-date mass-radius relation \citep{how01} expressed in
the form of
\begin{equation}
M_2=0.08 f^{-1.95}P_{\rm orb}^{1.3} \label{m2} 
\end{equation}
where $P_{\rm orb}$ is the binary orbital period in hours and $f$ is a
factor that describes the bloating of the secondary star in systems
above the period gap.  For the EX Hya orbital period of 1.6 hr and
$f=1$, we calculate $M_2=0.152\msun$, a value slightly higher than the
$0.128\msun$ that has been used previously to determine the white
dwarf mass.

Using the $K_1$ amplitude determined from our UV narrow emission line
radial velocity curve, we expect to obtain an unambiguous
determination of the mass of the white dwarf if our value of $M_2$ is
correct.  We employ a modified form of Kepler's third law,
\begin{equation}
\sin^3 i = \frac{K_1^3 P_{\rm orb}}{2 \pi G M_2}
\left(\frac{q+1}{q}\right)^2 \label{m1}
\end{equation}
and solve for $q$ in order to determine $\mwd$.  For
$i=78\pm1^{\circ}$ \citep{hellier87}, $M_2=0.152\msun$, and
$K_1=59.6\pm2.6\kms$, we determine a white dwarf mass,
$\mwd=1.33\pm0.11\msun$.  This value is considerably higher than
$0.78\pm0.17\msun$, which was determined by \citet{hellier87}, and the
even lower values of $\sim0.5\msun$ that have been determined from
X-ray studies (e.g., Fujimoto \& Ishida 1997).  Although our $K_1$
amplitude agrees within the errors with the value measured by
\citet{hellier87}, our white dwarf mass does not agree with theirs due
to the larger secondary mass used in the above calculation.

Smith, Cameron, \& Tucknott (1993) applied the method of skew-mapping
to detect the secondary stars in cataclysmic variables.  Applying this
to EX Hya, they found a velocity amplitude for the secondary motion of
$K_2=356\pm4\kms$.  This amplitude, along with our measured value of
$K_1$ and calculated value of $M_2$ may be used to apply another
determination of $M_1$ using $K_1/K_2 = M_2/M_1$.  This method gives a
white dwarf mass of $\mwd=0.91\pm0.05\msun$, which is $\sim30\%$ lower
than the value determined above, although still larger than the
earlier estimates (e.g., Hellier et al. 1987).

\subsection{Spectral Modeling}\label{model} 
Discerning exactly which components of the binary system are
contributing UV continuum flux (and at which spin phases) is an
important issue for thorough modeling of the {\it HST} spectra of EX
Hya.  Spin maximum phase would seem to be the most likely time in
which we would have a direct view of the white dwarf, but are there
other components also visible during this phase?  The accretion poles
(shock regions) are strong emitters in the X-ray and EUV
\citep{bel02}.  The irradiated material in the accretion curtain close
to the pole, however, may emit reprocessed X-ray and EUV radiation as
UV radiation, thus adding another UV emission component to the system.
The accretion disk, although truncated, is likely to add to the UV
continuum emission as well.

To model our {\it HST} spectra with a white dwarf photosphere, we must
determine which spectrum is the best representation of the white dwarf
by ascertaining which of the different components discussed above
contribute to the spin maximum and spin minimum spectra.  Visual
inspection of the spin maximum and minimum spectra reveals their
differences; the red-ward slopes of each spectrum are roughly the same
while the blue end of the spin minimum spectrum has a lower flux than
that of the spin maximum spectrum.  There are two possible cases to
consider; $1-$ spin minimum is the true white dwarf spectrum and the
accretion curtain is optically thin to the white dwarf continuum.
Spin maximum therefore contains an additional continuum component
that could be the accretion region located at either pole.  $2-$ The
spin minimum spectrum is absorbed by the accretion curtains and the
spin maximum spectrum is the true white dwarf spectrum plus an
additional continuum component.  These two options outline the
occultation (1) versus absorption (2) models.  For option 1, spin phased
modulation is due to occultation of an emitting pole, while option 2
dictates that spin phased modulation is due to UV flux being absorbed
by the accretion curtains.

\begin{figure}
\plotone{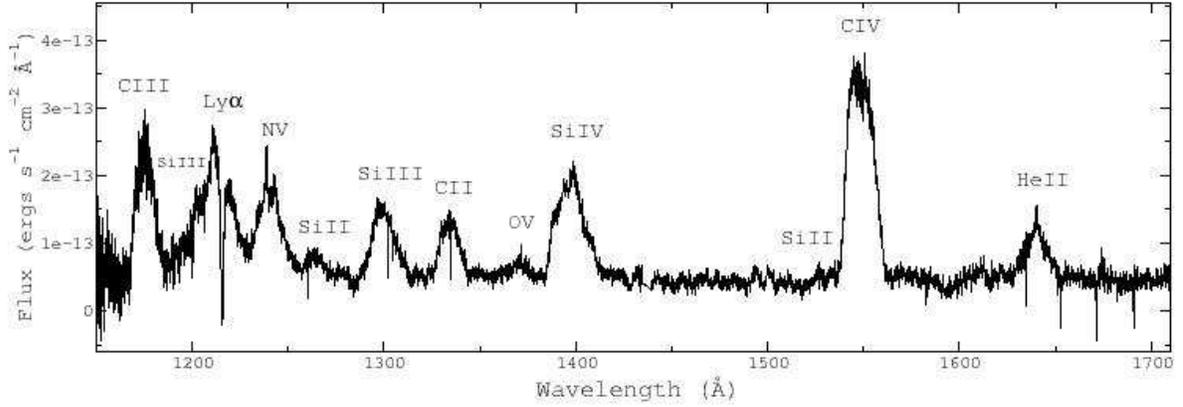}
\figcaption[]{Spin maximum - spin minimum difference spectrum. 
\label{sub_spec}}
\end{figure}

For either model option, the spin maximum spectrum has added UV
continuum components.  This is evidenced by the fact that the blue
slopes of the spin maximum and minimum spectra are not equal.  If the
spin modulation were simply due to absorption, the spin minimum
spectrum would retain the same slope as that of the spin maximum
spectrum.  (This assumes a gray absorber, which is valid over the
small, $\sim600$\AA, wavelength range covered by the {\it HST}
spectra.)  We show the difference spectrum of EX Hya, which is the
spin maximum minus the spin minimum spectrum, in Figure
\ref{sub_spec}.  This spectrum represents the additional UV continuum
and emission line components that are seen during spin maximum phases.
Considering the above arguments, we rule out the spin maximum spectrum
as representative of only the UV white dwarf continuum flux.  For
option 1 to be viable, the accretion curtains {\it near} the poles
would have to be dense enough to absorb the flux emitted from the
accretion region.  While the outer curtains are optically thin, it is
likely that the curtain close to the pole is optically thick and will
absorb emitted radiation from the accretion pole.  We therefore
suggest the case that the spin minimum spectrum is the better
representation of the pure white dwarf spectrum and that a blue
blackbody and emission line component (or multiple components) is seen
in addition to the white dwarf during spin maximum.  Flux contributed
by the truncated accretion disk must also be considered at all spin
phases.

\begin{figure}
\plotone{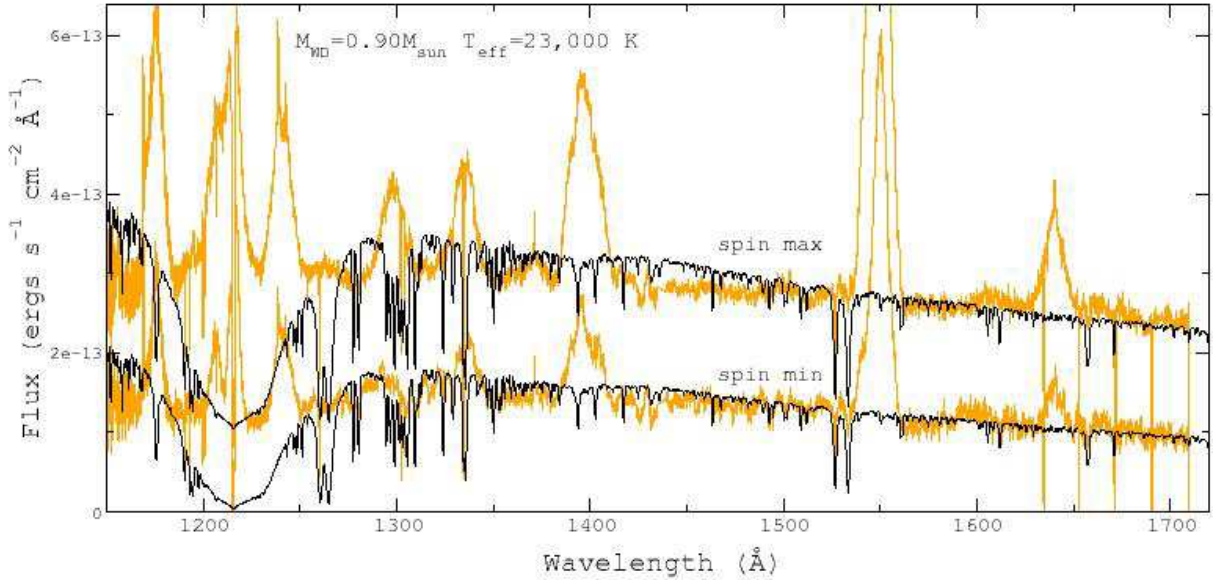} 
\figcaption[]{An $\mwd=0.90\msun$, $T_{\rm eff}=23000$ K white dwarf
photosphere model (black spectra) fit to the spin maximum (top) and
spin minimum (bottom) spectra (gray spectra).  The slope and
absorption lines of the spin minimum spectrum are fit better by the
white dwarf photosphere model than the spin maximum
spectrum. \label{fit}}
\end{figure}

We first created white dwarf photosphere models using TLUSTY and
SYNSPEC \citep{hub88} for the two white dwarf masses computed in
\S\ref{mass}, $0.90\msun$ ($\log\,g=8.5$) and $1.33\msun$
($\log\,g=9.4$).  We modeled white dwarf photospheres of solar
composition and considered temperatures ranging from $T_{\rm
eff}=19000\, {\rm K}\, - 35000$ K (in steps of $1000$ K).  Although it
was decided that the spin minimum spectrum would be the best
representation of the white dwarf spectrum, the model photospheres
were fit to the spin maximum and minimum spectra in order to insure
that the spin minimum spectrum was indeed the best choice.  For each
fit, the broad emission lines in the {\it HST} spectrum were masked,
except for the wavelength range including the
\ion{Si}{3}$\lambda1298$\AA\ and \ion{C}{2}$\lambda1335$\AA\ 
emission lines because of the white dwarf absorption components that
appear between these emission features.  Our best fitting $0.90\msun$,
$T_{\rm eff}=23000$ K white dwarf photosphere model is shown in Figure
\ref{fit} fit to both the spin maximum and spin minimum spectra.  The
spin minimum fit has a reduced $\chi^2$ of 1.7, and scaled to the {\it
HST} continuum fluxes, gives a distance to EX Hya of 54 pc.  The spin
maximum fit has a reduced $\chi^2$ value of 3.3 and scaled to the {\it
HST} continuum fluxes gives a distance of 46 pc.

Note that many of the absorption features appearing between the
\ion{Si}{3} and \ion{C}{2} in the spin minimum spectrum are fit well
by the white dwarf photosphere model.  The slope of the model spectrum
also produces a better match to that of the spin minimum spectrum than
the maximum spectrum.  These facts, coupled with the lower reduced
$\chi^2$ value of the spin minimum fit, corroborate with our earlier
statement that the spin minimum phases are the ideal phases during
which to model the white dwarf photosphere.  We therefore use only the
spin minimum spectrum in our following model fits.

White dwarf photospheres with temperatures of $T_{\rm
eff}=23000-35000$ K (in steps of $2000$ K) were created for
$\mwd=1.33\msun$.  The best fit model to the spin minimum spectrum has
a temperature of $T_{\rm eff}=25000$ K, but while this model provides
a good fit, it also places the distance of EX Hya at $30$ pc!  This
distance varies greatly from the distances of $65-100$ pc that have
been determined in previous studies \citep[e.g.,][]{hellier87,eis02}.
We explored the relationship of distance versus white dwarf
photosphere temperature for both white dwarf masses over a range of
temperatures and found that, for $\mwd=1.33\msun$, a white dwarf
temperature of $T_{\rm eff}=35000$ K would be required to achieve a
distance of 55 pc (when scaled to the spin minimum spectrum).  An
$\mwd=1.33\msun$, $T_{\rm eff}=35000$ K white dwarf photosphere model
was considered for the spin minimum spectrum, but the slope did not
match the {\it HST} continuum slope.  However, before we completely
rule out either white dwarf mass, we must also consider the accretion
disk contribution to the UV continuum flux.

Accretion disk models were created using TLUSDISK, SYNSPEC, and
DISKSYN \citep{hub88,wad98} for white dwarf masses of $\mwd=0.90\msun$
and $\mwd=1.33\msun$.  We used an accretion rate of $10^{16}{\rm g\,
s}^{-1}$ \citep{fuj97} and an inclination of $i=78^{\circ}$
\citep{hellier87}.  In order to determine the radius at which the disk
should be truncated, the inner disk radius was determined from our
optical spectra.  For a Keplerian disk, the highest velocities in an
emission line originate from the inner disk.  The optical line wings,
therefore, should map to the inner edge of the disk.  The H$\alpha$
$\lambda6563$\AA\ line profile in each of our optical spectra was used
to determine the velocities at the blue and red wings.  For a
$\mwd=0.90\msun$ white dwarf, these velocities map to an inner
accretion disk radius of $R_{\rm inner}=2.5\rwd$ ($1.6\times10^9\cm$)
and for a $\mwd=1.33\msun$ white dwarf, the inner radius becomes
$R_{\rm inner}=9.0\rwd$ ($2.3\times10^9\cm$).  Previous modeling
studies \citep[e.g.,][]{gre97} have assumed $R_{\rm inner}\sim5\rwd$
($3.5\times10^9\cm$) and optical spectroscopy by \citet{hellier87}
gave $R_{\rm inner}=6\times10^9\cm$ for $\mwd\sim0.8\msun$.

Two different accretion disk models must be created for the two
different white dwarf masses we have been using.  $\mwd=0.90\msun$ and
$\dot{M}=10^{16}\,{\rm g\,s}^{-1}$ correspond to Models $n$ and $p$ of
\citet{wad98} and $\mwd=1.33\msun$ with the same accretion rate
matches most closely with Models $y$ and $z$.  Model $n$ ($p$) has UV
flux contributing annuli that extend to $6.2\rwd$ ($9.40\rwd$), and
Model $y$ ($z$) has contributing annuli that extend out to $14.3\rwd$
($21.7\rwd$); the last annulus in each model has $T\sim10000$ K.
Although it would seem that the synthetic spectra for each disk model
would be different because of the different truncation points, the
truncated accretion disk spectra are in fact quite similar.  This is
due to the fact that the more massive white dwarf has a hotter
accretion disk and while the disk is truncated at a greater radius
than the less massive white dwarf, an approximately equal amount of UV
flux is still emitted from the remaining annuli.  

\begin{figure}
\plotone{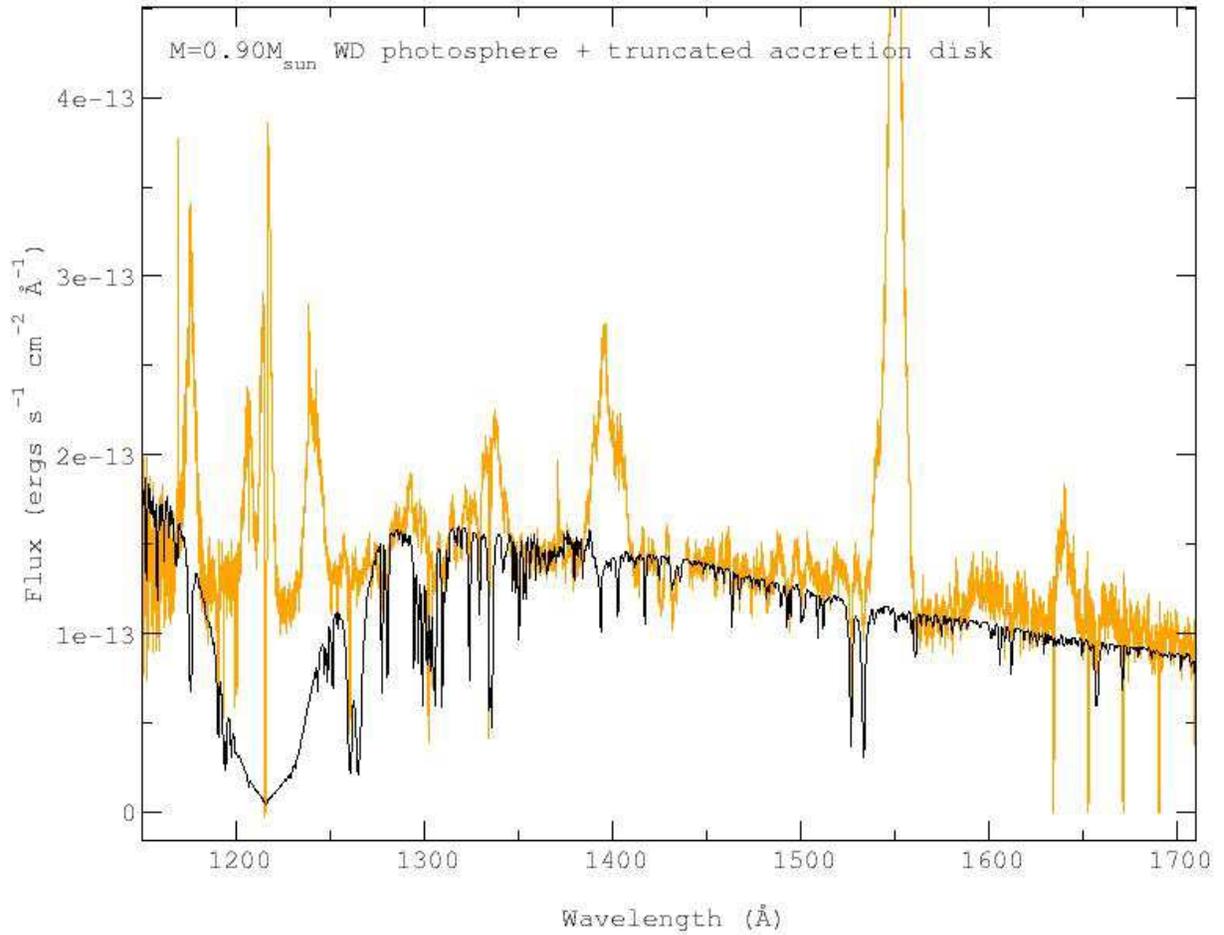} 
\figcaption[]{Best fit white dwarf photosphere plus truncated
accretion disk model for EX Hya.  This model uses $\mwd=0.90\msun$,
$T=23000$ K, and $R_{\rm inner}=2.5\rwd$, and is fit to the spin
minimum spectrum.  Scaled to the EX Hya spectra, this model gives a
distance of 60 pc.\label{fit3}}
\end{figure}

The truncated disk models were added to the white dwarf photosphere
models using the following method: the photosphere model and disk
model were scaled separately to a distance, $d$, and then summed
together.  The summed spectrum was then fit to the observed spectrum.
The distance, $d$, was modified until the summed spectrum produced a
good fit to the observed spectrum.  The summed model fits were judged
by visual inspection, rather than by a formal fitting routine.  For
$\mwd=0.90\msun$, we found that a $T_{\rm eff}=23000$ K white dwarf
plus the truncated disk produced the best fit at a distance $d=60$ pc.
The $\mwd=1.33\msun$ model showed a best fit for $T_{\rm eff}=25000$ K
plus the truncated disk at a distance of $d=33$ pc.  The
$\mwd=0.90\msun$ fit is shown in Figure \ref{fit3}, as we definitively
rule out the higher mass model below.

The white dwarf photosphere plus truncated accretion disk model fits
to the {\it HST} fluxes give two different distances: the $0.90\msun$
white dwarf model gives $d=60$ pc, while the $1.33\msun$ white dwarf
gives $d=33$ pc.  We can check these distances independently by using
Bailey's method \citep{bai82}, which uses the $K-$band magnitude and
measured surface brightness values for K-M single stars to determine
a distance;
\begin{equation}
\log\,d=\frac{K}{5}+1- \frac{S_K}{5}+ \log\left(\frac{R_2}{R_{\odot}}\right)
\label{distance} 
\end{equation}
where $S_K$ is the surface brightness.  Using $K=11.705(28)$ from the
2MASS survey \citep{hoard02}, $S_K=4.5\pm0.5$ for an M3 main sequence
star \citep{dhi97}, $R_2=0.17R_{\odot}$, and the secondary $K-$band
flux contribution as $75\pm25\%$ \citep{dhi97}, we determine the
distance to EX Hya to be $d=54^{+12}_{-7}$ pc.  This effectively rules
out the $1.33\msun$, $T_{\rm eff}=25000$ K white dwarf, as the
distance determined by this model is 33 pc.

The greatest uncertainties in the distance calculation and the white
dwarf mass determination are the secondary star mass and spectral
type.  An M3 main sequence secondary star \citep[determined from IR
spectra,][]{dhi97} has a mass of $\sim0.3\msun$.  This is not
consistent with the mass determined using the ZAMS mass-period
relation (Equation \ref{m2}), which gave the secondary mass as
$0.15\msun$.  Returning to Equation \ref{m1}, we find that a secondary
mass of $0.12\msun$ would be required to obtain $\mwd=0.9\msun$.  A
more accurate secondary mass will be needed to aid future distance and
white dwarf mass determinations of EX Hya.

\section{CONCLUSION}
We have made an accurate measurement of the $K$ amplitude of the white
dwarf based on the narrow emission line radial velocity curve.  The
small velocities and high ionization temperatures required for
production of the \ion{N}{5} and \ion{O}{5} emission lines indicate
that they are formed close to the white dwarf, and therefore, the
$K_1=59.6\pm2.6\kms$ value that we have measured is an accurate
representation of the motion of the white dwarf within the binary
system.

Synthetic spectral models for a white dwarf photosphere and truncated
accretion disk were made using TLUSTY and SYNSPEC \citep{hub88}.
Based on spectral fits using a white dwarf photosphere model, we
determined that the spin minimum phases are the best representation of
the white dwarf spectrum, i.e., occultation of (one of) the emitting
poles is responsible for the continuum modulation seen over the white
dwarf spin phase.  The distance we derive based on spectral modeling,
$d\approx60$ pc, agrees well with two recent measurements;
\citet{eis02} determine $d=65\pm11$ pc using Bailey's method and a $K$
magnitude derived from their infrared spectra, while a recent parallax
measurement has placed the distance to EX Hya at $\sim60$ pc
(K. Beuermann, private communication).  This implies that our spectral
modeling parameters for a solar composition white dwarf photosphere
and truncated accretion disk - $\mwd=0.9\msun$, $T=23000$ K, and
$R_{\rm inner}=2.5\rwd$ - represent EX Hya well.

\acknowledgments 
We thank the director of the Hubble Space Telescope for allocating DDT
time for the EX Hya multi-wavelength project.  We also thank
K. Beuermann for use of his EX Hya parallax distance prior to
publication.  The anonymous referee is thanked for very constructive
comments.  Support for proposal 8807 was provided by NASA through a
grant from the Space Telescope Science Institute, which is operated by
the Association of Universities for Research in Astronomy, Inc., under
NASA contract NAS 5-26555.  SBH acknowledges partial support of this
work from HST Grant GO-08807.01-A.

\begin{deluxetable}{cccc}
\tablecolumns{4}
\tablewidth{0pc}
\tablecaption{{\it HST} spectra phase coverage.
\label{phase}}
\tablehead{
\colhead{{\it HST} Data set} & \colhead{HJD start-stop} & 
\colhead{Spin Phase} &  \colhead{Binary Phase}}
\startdata
o68301010 & $2451682.894 - 2451682.919$ & $0.67-0.22$   &  $0.32-0.69$ \\ 
o68301020 & $2451682.950 - 2451682.982$ & $0.88-0.55$   &  $0.14-0.60$ \\
o68301030 & $2451683.017 - 2451683.049$ & $0.32-0.99$   &  $0.13-0.59$ \\
o68302010 & $2451694.622 - 2451694.647$ & $0.62-0.17$   &  $0.19-0.56$ \\
o68302020 & $2451694.680 - 2451694.712$ & $0.87-0.55$   &  $0.04-0.50$ \\
o68302030 & $2451694.747 - 2451694.779$ & $0.32-0.99$   &  $0.03-0.48$ \\
\enddata
\end{deluxetable}

\begin{deluxetable}{ccccc}
\tablecolumns{5}
\tablewidth{0pc}
\tablecaption{Sinusoidal fits to the continuum fluxes.
\label{contsinfit}}
\tablehead{
\colhead{Continuum Region} & \colhead{$A$\tablenotemark{a}} & 
\colhead{$B$\tablenotemark{a}} & \colhead{$\phi_0$} & \colhead{$B/A$}}
\startdata
$\ll1352-1362$\AA & 
	$1.683\pm0.001$ & $0.381\pm0.001$ & $0.77\pm0.05$ & $23\%$\\
$\ll1440-1520$\AA &
	$1.554\pm0.001$ & $0.315\pm0.001$ & $0.76\pm0.05$ & $20\%$\\
$\ll1565-1620$\AA & 
	$1.375\pm0.001$ & $0.337\pm0.001$ & $0.78\pm0.05$ & $24\%$\\
\enddata
\tablenotetext{a}{$A$ and $B$ have units of $10^{-13}\,\fluxa$.}
\end{deluxetable}

\begin{deluxetable}{ccccc}
\tablecolumns{5}
\tablewidth{0pc}
\tablecaption{Sinusoidal fits to the broad emission line fluxes.
\label{fluxsinfit}}
\tablehead{
\colhead{Ion} & \colhead{$A$\tablenotemark{a}} & 
\colhead{$B$\tablenotemark{a}} & \colhead{$\phi_0$} & \colhead{$B/A$}}
\startdata
\ion{C}{3} & $1.897\pm0.004$ & $1.381\pm0.003$ & $0.81\pm0.05$ & $73\%$\\
\ion{N}{5} & $1.694\pm0.002$ & $0.864\pm0.001$ & $0.80\pm0.05$ & $51\%$\\
\ion{Si}{3}& $1.183\pm0.001$ & $0.618\pm0.001$ & $0.83\pm0.05$ & $52\%$\\
\ion{C}{2} & $1.288\pm0.001$ & $0.555\pm0.001$ & $0.81\pm0.05$ & $43\%$\\
\ion{Si}{4}& $2.895\pm0.002$ & $1.732\pm0.001$ & $0.81\pm0.05$ & $60\%$\\
\ion{C}{4} & $6.771\pm0.004$ & $3.490\pm0.002$ & $0.80\pm0.05$ & $52\%$\\
\ion{He}{2}& $0.963\pm0.001$ & $0.554\pm0.001$ & $0.82\pm0.05$ & $58\%$\\
\enddata
\tablenotetext{a}{$A$ and $B$ have units of $10^{-12}\,\flux$.}
\end{deluxetable}  

\begin{deluxetable}{ccccc}
\tablecolumns{5}
\tablewidth{0pc}
\tablecaption{Sinusoidal fits to the broad emission line equivalent
widths.
\label{eqwsinfit}}
\tablehead{
\colhead{Ion} & \colhead{$A$\tablenotemark{a}} & 
\colhead{$B$\tablenotemark{a}} & \colhead{$\phi_0$} & \colhead{$B/A$}}
\startdata
\ion{C}{3} & $10.84\pm0.02$ & $-5.35\pm0.01$ & $0.81\pm0.05$ & $49\%$\\
\ion{N}{5} & $10.44\pm0.01$ & $-1.738\pm0.001$ & $0.82\pm0.05$ & $17\%$\\
\ion{Si}{3}& $7.20\pm0.01$ & $-2.141\pm0.002$ & $0.90\pm0.05$ & $30\%$\\
\ion{Si}{4}& $17.06\pm0.01$ & $-6.638\pm0.003$ & $0.82\pm0.05$ & $39\%$\\
\ion{C}{4} & $47.03\pm0.03$ & $-12.37\pm0.01$ & $0.83\pm0.05$ & $26\%$\\
\ion{He}{2}& $7.39\pm0.01$ & $-2.078\pm0.002$ & $0.85\pm0.05$ & $28\%$\\
\enddata
\tablenotetext{a}{$A$ and $B$ have units of \AA.}
\end{deluxetable}  

\clearpage
\newpage

%Figure 1

%Figure 2
\begin{figure}
\plotone{f2.eps}
\figcaption[]{Continuum flux measured at $\lambda\lambda1352-1362$\AA\
plotted over spin phase and fit with a sinusoidal function of the form
$f(10^{-13}\fluxa)=A+B\sin 2\pi(\ps-\phi_0)$, where $A=1.683\pm0.001$,
$B=0.381\pm0.001$, and $\phi_0=0.76\pm0.05$. \label{spcont_flux}}
\end{figure}

%Figure 3

%Figure 4

%Figure 5

%Figure 6

%Figure 7

%Figure 8

%Figure 9

%Figure 10

%Figure 11

%Figure 12


\begin{thebibliography}{99}
%
\bibitem[Bailey(1982)]{bai82}
Bailey, J.
1982, MNRAS, 197, 31
%
\bibitem[Belle et al.(2002)]{bel02}
Belle, K. E., Howell, S. B., Sirk, M. M., \& Huber, M. E.
2002, \apj, 577, 359
%
\bibitem[Dhillon et al.(1997)]{dhi97}
Dhillon, V. S., Marsh, V. T., Duck. S. R., \& Rosen, S. R.
1997, MNRAS, 285, 95
%
\bibitem[Eisenbart et al.(2002)]{eis02}
Eisenbart, S., Beuermann, K., Reinsch, K., \& G\"{a}nsicke, B. T.
2002, \aap, 382, 984
%
\bibitem[Fujimoto \& Ishida(1997)]{fuj97}
Fujimoto, R. \& Ishida, M.
1997, \apj, 474, 774
%
\bibitem[Gilliland(1982)]{gil82}
Gilliland, R. L.
1982, \apj, 258, 576
%
\bibitem[Greeley et al.(1997)]{gre97}
Greeley, B. W., Blair, W. P., Long, K. S., \& Knigge, C.
1997, \apj, 488, 419
%
\bibitem[Hellier et al.(1987)]{hellier87}
Hellier, C., Mason, K. O., Rosen, S. R., \& C\'{o}rdova, F. A.
1987, \mnras, 228, 463
%
\bibitem[Hellier \& Sproats(1992)]{ephem}
Hellier, C. \& Sproats, L. N.
1992, IBVS, 3724
%
\bibitem[Hoard et al.(2002)]{hoard02}
Hoard, D. W., Wachter, S., Clark, L. L., \& Bowers, T. P.
2002, ApJ, 565, 511
%
\bibitem[Howell et al.(2001)]{how01}
Howell, S. B., Nelson, L. A., \& Rappaport, S.
2001, \apj, 550, 897
%
\bibitem[Hubeny(1988)]{hub88}
Hubeny, I.
1988, Comput. Phys. Comm., 52, 103 
%
\bibitem[King \& Wynn(1999)]{king99}
King, A. R. \& Wynn, G. A.
1999, \mnras, 310, 203
%
\bibitem[Mauche(1999)]{mauche99}
Mauche, C. W.
1999, \apj, 520, 822
%
\bibitem[Patterson(1984)]{pat84}
Patterson, J.
1984, \apjs, 54, 443
%
\bibitem[Rosen et al.(1991)]{rosen91}
Rosen, S. R., Mason, K. O., Mukai, K., \& Williams, O. R.
1991, \mnras, 249, 417
%
\bibitem[Smith et al.(1993)]{smith93}
Smith, R. C., Cameron, A. C., \& Tucknott, D. S.
1993, in Ann. Israel Phys. Soc., Vol 10, 
Cataclysmic Variables and Related Physics, eds. O. Regev \&
G. Shaviv, 70
%
\bibitem[Wade \& Hubeny(1998)]{wad98}
Wade, R. A. \& Hubeny, I.
1998, \apj, 509, 350
%
\bibitem[Warner(1976)]{war76}
Warner, B.
1976, in IAU Symp. No. 73, The Structure and Evolution of Close Binary
Systems, eds. P. Eggleton, S. Mitton, \& J. Whelan, (Dordrecht: Reidel), 85
%
\end{thebibliography}
\end{document}